\documentclass[aps,prd,a4paper,twocolumn,showpacs,showkeys,preprintnumbers,amsmath,amssymb,nofootinbib]{revtex4}
\usepackage{graphicx}
\usepackage{bm}
\usepackage{dcolumn}
\usepackage{color}

\newcommand{\hide}[1]{{}}
\newcommand{\mr}[1]{\mathrm{#1}}
\newcommand{\si}[1]{{\scriptscriptstyle{#1}}}

\newcommand{\ads}{\ensuremath{\mathrm{AdS}_{5}\,}}
\newcommand{\ZZ}{\ensuremath{\mathbb{Z}}}
\newcommand{\kappafive}{\kappa_{5}}
\newcommand{\kappareduct}{\kappa_{4}}
\newcommand{\tension}{\mathcal{T}}

\newcommand{\bb}{{\si{\bullet}}}
\newcommand{\ve}[1]{\boldsymbol{#1}}
\newcommand{\p}{\partial}
\newcommand{\ud}{\mathrm{d}}
\newcommand{\Hb}{h_{\bb}}

\newcommand{\vv}{{v}}

\newcommand{\de}{\delta}
\newcommand{\ga}{\gamma}

\newcommand{\dd}{\partial}
\newcommand{\ra}{\rightarrow}
\newcommand{\al}{\alpha}
\newcommand{\om}{\omega}
\newcommand{\lsim}{\,\raisebox{-0.6ex}{$\buildrel < \over \sim$}\,}
\newcommand{\gsim}{\,\raisebox{-0.6ex}{$\buildrel > \over \sim$}\,}
\newcommand{\be}{\begin{equation}}
\newcommand{\ee}{\end{equation}}
\newcommand{\ben}{\begin{equation*}}
\newcommand{\een}{\end{equation*}}
\newcommand{\bea}{\begin{eqnarray}}
\newcommand{\eea}{\end{eqnarray}}
\newcommand{\bean}{\begin{eqnarray*}}
\newcommand{\eean}{\end{eqnarray*}}
\newcommand{\tin}{t_{\rm in}}

\begin{document}

\title{Graviton production in anti-de Sitter braneworld cosmology: \\
         A fully consistent treatment of the boundary condition}

\author{Marcus Ruser, Ruth Durrer, 
       Marc Vonlanthen and Peter Wittwer}
\affiliation{Universit\'e de Gen\`eve, D\'epartment de Physique Th\'eorique, 
        24 quai Ernest Ansermet, CH--1211 Gen\`eve 4, Switzerland}

\date{\today}

\begin{abstract}
In recent work by two of us, [Durrer \& Ruser, PRL  {\bf 99}, 071601 (2007); 
Ruser \& Durrer PRD {\bf 76}, 104014 (2007)], graviton production due to a moving 
spacetime boundary (braneworld) in a five dimensional bulk has been considered. 
In the same way as the presence of a conducting plate modifies the electromagnetic 
vacuum, the presence of a brane modifies the graviton vacuum. As the brane moves, the 
time dependence of the resulting boundary condition leads to particle creation via the so called 'dynamical Casimir effect'. In our previous work a term in the boundary 
condition which is linear in the brane velocity has been neglected. In this work we 
develop a new approach 
which overcomes this approximation. We show that the previous results are 
not modified if the  brane velocity is low.
\end{abstract}


\maketitle

\section{Introduction}
The idea that our Universe is a $3+1$ dimensional membrane in a higher 
dimensional 'bulk' spacetime has opened new exciting prospects for 
cosmology, for reviews see~\cite{roy,mine}. In the simplest braneworlds 
motivated by string theory, the standard model particles are confined to 
the brane and only the graviton can propagate in the bulk. Of particular 
interest is the Randall-Sundrum (RS) model~\cite{RS1,RS2}, where the bulk 
is  5-dimensional anti-de Sitter space. If the so called RS fine tuning 
condition is satisfied, it can be shown that gravity on the brane 
'looks 4-dimensional' at low energies.

Within this model, cosmological evolution can be interpreted as the 
motion of the physical brane, i.e. our Universe, through the 5d bulk, 
 acting as a moving boundary for bulk fields, in particular 
for 5d gravitational perturbations.
Such a time-dependent boundary does in general lead to particle 
production via the dynamical Casimir effect~\cite{bordag, dodonov}.

Of course one can always choose coordinates with respect to which the brane 
is at rest, e.g. Gaussian normal coordinates. 
This leads to a time dependent bulk resulting in the same effect, 
particle production from vacuum due to a time varying background metric.
But then, usually (except in the case of de Sitter expansion on the 
brane~\cite{ruba}), the perturbation 
equation describing the evolution of gravitons is not separable and can only be 
treated with numerical simulations~\cite{jap1,kazu,seahra}. 
Furthermore, in a time dependent bulk,
a mode decomposition is in general ambiguous and one cannot split 
the field in a zero mode and Kaluza-Klein (KK) modes in a unique way.
One of the advantages of the dynamical Casimir effect approach presented 
in \cite{DR,RD} is that it allows for a clear physical interpretation and 
in addition exhibits an analogy with quantum electrodynamics.

\begin{figure}[ht]
\includegraphics[width=8cm]{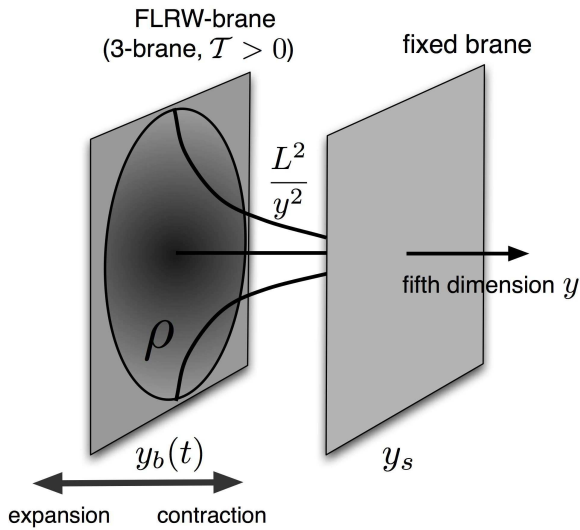}
\caption{\label{f:1} Two branes in an AdS$_5$ spacetime. The physical
  brane, a Friedmann universe with energy 
density $\rho$ is on  the left. 
  While it is approaching the static brane its scale 
  factor is decreasing, the Universe is contracting, and when it moves away 
  from the static brane the Universe is expanding. $L$ is the 
  AdS curvature radius which is related to the brane tension ${\cal T}$ via 
  Eq.~(\ref{e:fine}). The value of the scale factor of the brane metric as 
  a function of the extra dimension $y$ is also indicated.}
\end{figure}

Based on the picture of a moving brane in AdS$_5$, we have studied graviton 
production in an ekpyrotic type scenario~\cite{ekpy} where our Universe 
first approaches a second static brane. After a 'collision' the physical 
brane reverses direction and moves away from the static brane, see 
Fig.~\ref{f:1}. For an observer 
on the brane, the first phase corresponds to a contracting Universe, the 
collision represents the 'Big Bang' after which the Universe starts expanding
(see Fig.~1). We do not model the details of this collision, but
assume that the brane distance is still finite at the collision. This 
corresponds to a cutoff of all the physics which happens at scales smaller 
than the minimal brane distance when contraction reverses into expansion. In 
our results we assume this to be of the order of the string scale. We cut off
the spectra at the string scale. This is a conservative assumption which 
signifies that we neglect all the particle creation at energies higher than 
this scale.

We have obtained the following important results in our previous 
papers~\cite{DR,RD}: first of all, the energy density of KK 
gravitons in AdS$_5$ scales like stiff matter, $\propto a^{-6}$, where $a$
denotes the scale factor introduced in Eq.~(\ref{eq:branemetric}). 
Therefore, KK  gravitons in AdS$_5$ cannot represent the dark matter in 
the Universe
\footnote{See \cite{ruth} for a discussion on a contradicting 
result in the literature.}.
We have also seen that in the early Universe the back reaction 
from KK gravitons on the bulk geometry is likely to be important. Finally, we 
have derived a limit for the maximal brane velocity, the bounce 
velocity, $v_{\max}\lsim 0.2$ in order 
not to over-produce zero-mode (i.e. 4d) gravitons, 
the energy density of which is constrained by 
the nucleosynthesis bound. We have also calculated the spectra of both, 
the zero-mode and the KK gravitons.

In this previous work we have, however, neglected a term linear in the 
brane velocity $v$ in the boundary conditions (junction conditions) for 
the tensor perturbations. Here we derive a method which includes this term 
and allows to treat the problem without any low velocity approximation. 
We show that the low velocity results previously obtained 
are not modified. Especially, the nucleosynthesis bound on the maximal brane
velocity, $v_{\max}\lsim 0.2$, remains valid. In a 
subsequent study we shall investigate graviton production from
branes which achieve high velocities in detail~\cite{pap2}.

The paper is organized as follows. In the next section we repeat 
the basic equations for the evolution of tensor perturbations (gravitons)
and we explain why it is not straight forward to 
include the velocity term of the boundary condition. In 
Section~\ref{s:time-dep} we  present the new approach and obtain the modified 
perturbation equations via a coordinate transformation which is such that the 
velocity term in the boundary condition disappears. We then  quantize the 
system in the new coordinates. In Section~\ref{s:num}
we show 
numerical results for graviton production at relatively low velocities. In 
Section~\ref{s:con} we conclude. Technical details are deferred to
appendices.

\section{A moving brane in $\ads$}
%
\subsection{The background}
In Poincar\'e coordinates $(x^A)=(t, {\bf x}, y)$ with ${\bf x} = (x^1, x^2, 
x^3)$ and $A=0,...,4$,  the AdS$_5$ (bulk) metric is given by
\begin{equation}\label{e:bulk-metric}
 ds^2
 = g_{\si{AB}} d x^{\si{A}} d  x^{\si{B}}
 = \frac{L^2}{y^2} \left[-d t^2 + \delta_{ij} d x^i d x^j + dy^2\right]~,
\end{equation}
where $i,j=1,2,3$ and $L$ is the AdS$_5$ curvature radius  which is related
to the bulk cosmological constant by the 5d Einstein equation, 
$-\Lambda=6/L^2$. The physical brane representing our (spatially flat) 
Universe is located at some time dependent
position $y=y_b(t)$ in the bulk, and the metric induced on the brane 
is the Friedman-Robertson-Walker metric
\begin{eqnarray}
 ds^2 = a^2(\eta)\left[ -\ud\eta^2 +\delta_{ij}d x^id x^j\right]~,
 \label{eq:branemetric}
\end{eqnarray}
with scale factor $a(\eta)$ which is given by the brane position,
\begin{equation}
a(\eta)=\frac{L}{y_b(t)}~.
\label{e:a of y}
\end{equation}
The conformal time  $\eta$ of an observer on the brane,
 is related to the bulk time $t$ via
\begin{equation}
 d\eta = \sqrt{1- v^2}dt \equiv \ga^{-1}dt~.
\end{equation}
Here we have introduced the brane velocity
\begin{eqnarray}
 v \equiv \frac{dy_b}{dt} = -\frac{LH}{\sqrt{1+L^2H^2}}\quad \mbox{
 and }~  \ga  = \frac{1}{\sqrt{1-v^2}} ~. \label{e:vel}
\end{eqnarray}
$H$ is the usual Hubble parameter,
\begin{equation}
   H \equiv \frac{1}{a^2}\frac{da}{d\eta} \equiv 
   a^{-1}{\cal H}= -L^{-1}\ga v~.
\end{equation}
Its dynamics, as a result of the second junction condition, is determined 
by the modified Friedmann equation~\cite{roy}
\begin{equation}
H^2 = \frac{\kappa_4 \rho}{3} \left( 1+ \frac{\rho}{2{\cal T}}\right)\,,
\end{equation}
where ${\cal T}$ is the brane tension, $\rho$ the energy density on the 
brane, and we assume the RS fine tuning condition~\cite{RS1}
 \begin{equation}
 \frac{\kappafive^2 \tension^2}{12} = \frac{3}{L^2}~.
 \label{e:fine}
\end{equation}
Furthermore (see~\cite{RS1}), 
\begin{equation}
 \kappareduct =8\pi G_4 = \frac{\kappafive^2 \tension}{6}~.
\end{equation}
 We define the string and Planck scales by 
\begin{equation}
\kappa_5  =\frac{1}{M_5^3} = L_s^3~, \qquad
\kappa_4 = \frac{1}{M_{\rm Pl}^2}= L_{\rm Pl}^2~.
\label{e:string and Pl scale}
\end{equation}
Note that the RS fine-tuning condition is equivalent to
\begin{equation}
\kappa_5=\kappa_4\,L~ \mbox{ or }~~ \frac{L_s}{L}=\frac{L_{\rm Pl}^2}{L_s^2}. 
\label{e:RS fine tuning 0}
\end{equation}
Identifying $\kappa_5$ with the string scale is based on the 
assumption that this phenomenological model comes from string theory with 
one large extra-dimension $L$, the $y$ direction, while all the other 
extra-dimensions remain of the order of the string scale, $L_s$. In this 
case the 4d observed Planck scale is related to the string scale by 
Eq.~(\ref{e:RS fine tuning 0}).

\subsection{The setup}\label{s:setup}
%
We consider a radiation dominated brane which moves frome the Cauchy 
horizon, $y=0$, at $t=-\infty$ to a position $y_b(0)<y_s$ at $t=0$, where it 
bounces and changes its direction. In a radiation dominated universe 
$\rho\propto a^{-4}\propto y_b(t)^4$. Defining
\be
r(t) \equiv \frac{L_{\rm Pl}^2L^2\rho}{3}\ ,
\ee
we have $HL = \sqrt{r(1+r/4)}$. Inserting this in Eq.~(\ref{e:vel}) yields
\be \label{e:ydot}
\dot{y}_b(t) = v(t) = \pm\frac{\sqrt{r(t)(1+r(t)/4)}}{1+r(t)/2} \ .
\ee
Here the upper sign is chosen for negative times, when $y$ is growing and
the universe is contracting while the lower sign corresponds to positive 
times (expanding universe). At the bounce the maximal velocity, $v(0)$ is 
reached corresponding to the maximal radiation density given by
\be
r(0) = \frac{1}{2}\left(\sqrt{1+\frac{v^2(0)}{1-v^2(0)}} - 1\right)
\ee
At $t\neq 0$ the radiation density is 
$$ r(t) = r(0)\left(\frac{y_b(t)}{y_b(0)}\right)^4 \ . $$
Note that since the differential equation (\ref{e:ydot}) is first order, only 
one initial condition, e.g $v(0)$ can be chosen. $y_b(0)$ is then determined 
by the implicit equation 
$$
y(0)=\int_{-\infty}^0 v(t)dt \ .
$$
(Implicit because it contains $y(0)$ also in the integrand.)
Because of this complication it is simpler to choose the initial conditions 
at some early time, $\tin\ll 0$ so that $r(\tin)\ll1$. For $t\lsim \tin$
we can then approximate Eq.~(\ref{e:ydot}) to
$$ \dot{y}_b(t) = \sqrt{r(t)} = \sqrt{r(\tin)}\frac{y_b^2(t)}{y_b^2(\tin)} $$
with solution 
\be y_b(t) = -\frac{\sqrt{r(\tin)}\tin^2}{t} \ , \qquad t\le \tin \ . \ee
The initial condition $(\tin,r(\tin))$ determines the bounce velocity $v(0)$.

In this first paper, where we mainly want to present 
the method how to transform mixed boundary conditions into Neumann boundary
conditions, we simplify the background evolution by assuming $HL\ll 1$ or,
equivalently,  $r(t)\ll 1$ at all times. This is of course not a good 
approximation if the bounce velocity 
is high and we shall treat the brane motion correctly in Ref.~\cite{pap2}. 
With this (\ref{e:ydot}) reduces to
\be\label{e:low} \dot y_b(t) = \sqrt{r(t)} \propto y_b^2(t) \ee
at all times.  The expression
\be\label{e:ytlow}
 y_b(t) = \frac{L^2}{|t| +t_b} \ ,
\ee
with parameter $t_b$ solves Eq.~(\ref{e:low}) for all $t\neq 0$. Furthermore, 
it has the correct asymptotics and the bounce velocity is given by
$$ v(0) = \frac{L^2}{t_b^2} =\frac{y_b(0)^2}{L^2} \equiv v_b\ . $$
The kink at $t=0$ can be regularized by replacing $|t|$ by 
$\sqrt{t^2 +t_c^2}$, where $t_c$ is a small regularization parameter. For
$|t| \ll t_c$ this does not affect the dynamics but for $|t|<t_c$ the velocity 
is reduced and it actually passes through zero at $t=0$. For graviton 
frequencies with $\omega t_c\ll 1$ the particle production obtained is 
independent of this regularization.

\subsection{Tensor perturbations}
%
Allowing for tensor perturbations $h_{ij}(t,{\bf x},y)$ of the spatial 
three-dimensional geometry at fixed $y$, the perturbed bulk metric reads
\begin{align}
 \ud s^2 = \frac{L^2}{y^2}
 \left[-\ud t^2+(\delta_{ij}+2h_{ij})\ud x^i \ud x^j+d y^2 \right]~.
\end{align}
Tensor modes satisfy the traceless and transverse conditions, $h_i^i =
\p_ih^i_j = 0$. 
These conditions imply that $h_{ij}$ has only two independent degrees 
of freedom, the two polarization states $\bullet=\times,+$.
We decompose $h_{ij}$ into spatial Fourier modes,
\begin{equation}
 h_{ij}(t,\ve{x},y)
 = \int \frac{d^3k}{(2\pi)^{3/2}} \sum_{\bb=+,\times}
  e^{i\ve{k}\cdot\ve{x}}e_{ij}^{\bb}({\bf k})\Hb(t,y;{\bf k})~,
\label{e:h fourier decomposition}
\end{equation}
where $e_{ij}^{\bb}({\bf k})$ are unitary constant transverse-traceless
polarization tensors which form a basis of the two polarization
states $\bullet = \times,+$. Since the problem at hand obeys parity 
symmetry, we shall neglect in the
following the distinction between the two graviton polarizations and
consider only one of them. We then have to multiply the final results
for e.g. particle number or energy density by a factor of two to account 
for both polarizations.

Here we only consider 4d gravitational waves. The 5d metric has in principle
five different spin-2 polarizations. Two of them are the 
ones discussed here. In addition there are the two helicities of 
the so-called gravi-vector and a gravi-scalar (see, e.g.~\cite{CR}).
The gravi-vector and the gravi-scalar obey exactly the same propagation 
equation as the 4d gravitational waves in the bulk, only their boundary 
conditions are different. In 
principle they would add to the results obtained here. In this sense our 
results are conservative, but since the different polarization states do not
interact at the linear level they can be calculated independently. 
 These polarizations are expected to contribute on the same level as the two
considered here.

The perturbed Einstein equations and the second junction condition 
lead to the boundary value problem 
\begin{equation}
 \left[\p_t^2 +k^2 -\p_y^2 + \frac{3}{y}\p_y \right] h(t,y;{\bf k}) = 0 ~~ \mbox{in the bulk} 
\label{e:T-bulk-eq}
\end{equation}
and
\begin{equation}
        \left . \ga\left(\vv\dd_t +\dd_y\right)h \right|_{y_b(t)}=0 
\label{e:T-JC-simple}
\end{equation}
describing the time-evolution of the tensor perturbations as the brane moves
through the bulk. We introduce also a second, static brane at position $y_s$, 
which requires the additional boundary condition
\begin{equation}
\left . \dd_yh \right|_{y_s}=0~. 
\label{e:T-JC-stat}
\end{equation}
Eq.~(\ref{e:T-bulk-eq}) is the Klein-Gordon equation for a minimally coupled
massless mode in $\ads$, i.e. the operator acting on $h$ is just the 
Klein-Gordon operator 
\begin{equation}
          \Box = \frac{1}{\sqrt{-g}} \partial_A 
          \left[ \sqrt{-g} g^{AB} \partial_B\right]~.
\label{e:KG operator}
\end{equation}

Equation (\ref{e:T-JC-simple})  is a time-dependent boundary 
condition (BC) coming from the fact that the moving brane acts like a "moving 
mirror" for the gravitational perturbations. Only in the rest-frame of the brane
do we have pure Neumann BCs. In a generic frame we have the 
Lorentz transformed BC which contains a velocity term $v\dd_t$.

We assume that the brane is filled with a perfect fluid
such that there are no anisotropic stress perturbations in the 
brane energy momentum tensor, i.e. there is no coupling of gravitational waves 
to matter. If this were the case, the r.h.s. of Eq.~(\ref{e:T-JC-simple}) would 
not be zero but a term coupling $h_{ij}$ to the matter on the brane,
see Eq.~(2.25) of \cite{RD},  would be present. 

 The analogy to a moving mirror is actually not just a pictorial one. 
Transverse-magnetic modes of the electromagnetic field in an 
 ideal, i.e. perfectly conducting, dynamical cavity are subject to the very 
same boundary condition, see, e.g., \cite{Diego}. In this context, the 
boundary condition (\ref{e:T-JC-simple}) is sometimes referred 
to as "generalized Neumann" boundary condition,  a terminology which we also 
adopt here. If the cavity is non-perfect, then also in the case of the 
electromagnetic field, the right hand side of the boundary condition
contains a term describing the interaction of the photon field with cavity 
material, similar to the anisotropic stress perturbations for the gravitational 
case considered here. 
This suggests that a brane with no anisotropic stresses could be termed 
"ideal brane". 

For the tensor perturbations the gravitational action up to
second order in the perturbations reads
\bea
{\cal S}_h &=& 4\,\frac{L^3}{2\kappa_5} \int dt\! \int d^3k 
\!\int_{y_b(t)}^{y_s} \frac{dy}{y^3}  \Big[|\partial_t h|^2 
- |\partial_y h|^2 \nonumber \\
 &&\qquad   \qquad\qquad\qquad \qquad\qquad -k^2|h|^2 \Big]~.
\label{e:action h}
\eea
One factor of two in the action is due to ${\mathbb Z}_2$ symmetry while
a second factor comes from the two polarizations.
As we have shown in~\cite{RD},  the BC's (\ref{e:T-JC-simple},\ref{e:T-JC-stat})
are indeed the only ones for which $\delta {\cal S}_h = 0$ leads to the free 
wave equation (\ref{e:T-bulk-eq}). (In principle also Dirichlet BC's,  i.e. 
$h$ vanishing identically on the brane, lead to a wave equation in 
the bulk. But besides leaving no room for a non-trivial dynamics of the 
gravitational waves on the brane, these are not obtained from the Einstein
equations in the bulk.)
%
\subsection{Dynamical Casimir effect approach}
%
 The wave equation (\ref{e:T-bulk-eq}) itself is not time dependent and
simply describes the propagation of free modes. It is
the time dependence of the BC (\ref{e:T-JC-simple}) that sources the
non-trivial time-evolution of the perturbations. 
As it is well known, such a system of a wave equation and a time-dependent
BC leads, within a quantum mechanical formulation, to particle production
from vacuum fluctuations. 
In the context of the photon field perturbed by a moving mirror this 
goes under the name ``dynamical Casimir 
effect''~\cite{bordag,dodonov}.

In \cite{DR,RD} we have extended a formalism which has been successfully 
employed for the numerical investigation of photon production in dynamical 
cavities \cite{Ruser:2004,Ruser:2005xg,Ruser:2006xg} to the RS braneworld 
scenario. We have studied graviton production by a moving brane,  which we 
call dynamical Casimir effect for gravitons, for a bouncing braneworld scenario.

However, in order to solve the problem, we have neglected the velocity term 
in the BC~(\ref{e:T-JC-simple}). The ansatz 
$$ h= \sum_{\al =0}^\infty a_\al(t)
    e^{-i\om_{\al,k} t}\phi_\al(t,y) + {\rm h.c.}~, ~~ 
  \om^2_{\al,k} =k^2 +m^2_\al(t)$$
then leads to a Sturm--Liouville problem for the instantaneous 
eigenfunctions $\phi_\al(t,y)$ consisting of the differential equation
\begin{equation}
\left[ -\partial_y^2 + \frac{3}{y} \partial_y \right] \phi_\alpha(t,y) 
= m_\alpha^2(t) \phi_\alpha(t,y)
\label{e:mode}
\end{equation}
and Neumann BC's at both branes. 
The  solutions of (\ref{e:mode}) respecting Neumann BC's at both branes are
\bea
\phi_0(t) &=& \frac{y_s y_b(t)}{\sqrt{y_s^2 - y_b^2(t)}}
\label{zero mode phi}\\
\phi_n(t,y) &=& N_n (t) y^2C_2(m_n(t),y_b(t),y) \nonumber \\  
~ \mbox{ with} && 
\nonumber \\   \hspace*{-3mm} \label{massive mode phi}
C_\nu(m,x,y) &=& Y_1(m x) J_\nu(my)\! -\! J_1(m x) Y_\nu(m y)\,.
\eea
They form a complete orthonormal system with respect to the inner product
\begin{equation}
(\phi_\alpha, \phi_\beta) = 2 \int_{y_b(t)}^{y_s} \frac{dy}{y^3} 
 \phi_\alpha(t,y) \phi_\beta(t,y) = \delta_{\alpha\beta} 
\label{e:orthonormal}
\end{equation}
and the completeness relation implies
\begin{equation}
2 \sum_{\alpha}  \phi_\alpha(t,y) \phi_\alpha(t,\tilde{y}) = 
  \delta(y - \tilde{y}) y^3~.
\label{e:complete}
\end{equation}
The factor two accounts for the $\ZZ_2$ symmetry of the bulk.

In \cite{RD} we call $\phi_0$ and $\phi_n$ the zero-mode and
Kaluza-Klein (KK)- mode solution, respectively. 
Here $\phi_0$ is the massless mode, $m_0=0$, which reduces to the
usual $3+1$ - dimensional graviton on the brane. The KK masses 
$m_n\neq 0$ are determined by the BC at the static brane, see, 
e.g.~\cite{RD,CDR} for more details.

Due to the completeness and ortho-normality of the functions $\{\phi_\alpha\}$
at any instant in time, any general solution of (\ref{e:mode}) subject to 
Neumann BC's can be expanded in these instantaneous eigenfunctions. 
If we add the term $v\dd_t$ to the boundary condition this feature is lost and
we can no longer expect to find a complete set of instantaneous eigenfunctions.

However, since the entire effect disappears when the velocity tends to zero,
neglecting a term which is first order in the velocity seems not to be a 
consistent approach. This problem prompted us to search for another 
description allowing us to treat the boundary condition 
(\ref{e:T-JC-simple}) in full.

%
\section{Graviton production in a time-dependent bulk with a moving brane}
\label{s:time-dep}
%
In this section we introduce a new time coordinate 
which is chosen such that the velocity term in the boundary condition
disappears but the mode equation for the instantaneous eigenfunctions
still remains the Bessel equation (\ref{e:mode}) with its solution given by 
Eqs.~(\ref{zero mode phi}) and (\ref{massive mode phi}). We then extend 
the formalism of \cite{RD} to this case and shall see that for small 
velocities our previous results are not modified.

\subsection{A new time coordinate}
We introduce new variables $(\tilde{x}^A) = (\tau,{\bf x},z)$ given by
\begin{equation}
\tau(t,y) = t + s(t,y)~,~~z = y~. 
\label{e:trafo}
\end{equation}
The idea is to find a function $s(t,y)$ such that 
$\tau \ra t$ for all $y$, when $v\ra 0$ and
that the junction condition (\ref{e:T-JC-simple}) reduces to a normal 
Neumann BC in the new variables.  We can then use the mode functions 
(\ref{zero mode phi}) and (\ref{massive mode phi}) to formulate the 
problem quantum mechanically.
One might first be tempted to make a $y$-dependent Lorentz transformation 
to the rest frame of the moving brane, but actually this does not lead to
Neumann  BC's in our case as the transformation induces 
new terms in the metric. We therefore first leave the function $s(t,y)$
completely general and formulate the conditions which have to be satisfied
in order for the new BC's to be purely Neumann.

In $(\tau,{\bf x}, z)$-coordinates, the brane trajectory is given by 
the implicit equation 
 \begin{equation}
 z_b(\tau) = y_b\big[t(\tau,z_b(\tau))]~.
 \label{e:zbrane motion}
 \end{equation}
Once we have specified the function $s(t,y)$, the new brane
trajectory $z_b(\tau)$ can be found. This is done numerically since neither 
$s(t,y)$ nor the inverse $t(\tau,z)$ of (\ref{e:trafo}) exist in closed form. 
As in \cite{RD} we restrict ourselves to brane motions where asymptotically, 
i.e. for $t \rightarrow \pm \infty$, the physical brane approaches the Cauchy 
horizon ($y_b\rightarrow 0$), moving very slowly $(v\rightarrow 0)$.

The new metric given by
\begin{equation}
ds^2 = \tilde g_{AB}(\tau,z)d\tilde x^A\tilde x^B
\label{e:new metric}
\end{equation}
is time dependent and contains non-vanishing cross terms $ \tilde g_{0z}$.
The explicit expression is given in (\ref{e:new metric expl}).
We now show that the function $s(t,y)$ can be chosen such that the 
time-derivative term in the boundary condition disappears.

In the coordinates defined in Eq.~(\ref{e:trafo}), the junction condition 
(\ref{e:T-JC-simple}) becomes
\bea
\left[ v(t)  \frac{\partial \tau}{\partial t} +  \frac{\partial\tau}{\partial y}  \right]\partial_\tau h(\tau,z)
+ \partial_z h(\tau,z) = && \nonumber \\
 \left[ v(t)  \left\{1 + \partial_t s(t,y)\right\} +  \partial_y s(t,y) \right]\partial_\tau h(\tau,z) && \nonumber \\
 \qquad + \partial_z h(\tau,z) = 0 \quad \mbox{ at } z=z_b(\tau)~.&&
\eea
In order to obtain Neumann boundary conditions, we require that the term 
in square brackets vanish at $z_b(\tau)$. This 
leads to the condition 
\begin{equation}
- \left.\frac{\partial_y s(t,y)}{1 + \partial_t s(t,y)} \right|_{y = y_b(t)} 
   = v(t) 
\label{e:s condition 1}
\end{equation}
for the function $s(t,y)$. Furthermore, we want to maintain the Neumann BC 
at the static brane $y_s = z_s$. This yields the additional condition for 
the function $s(t,y)$
\begin{equation}
\partial_y s(t,y)|_{y = y_s} = 0~.  
\label{e:s condition 2}
\end{equation}
Hence, if we can find a function $s(t,y)$ which satisfies Eqs. 
(\ref{e:s condition 1}) and (\ref{e:s condition 2}), the junction conditions 
in the new coordinates reduce to  Neumann BC's
\begin{equation}
\partial_z h(\tau,z) = 0~{\rm at}~~ z = z_s ~~ {\rm and} ~~ z= z_b(\tau)~.
\end{equation}

To find a suitable function $s(t,y)$ we choose the separation ansatz
\begin{equation}\label{e:seqfsi}
s(t,y) = f (t) \sigma(y)
\end{equation}
leading to 
\begin{equation} \label{vtft}
v(t) = - \frac{f(t) \partial_y \sigma\big(y_b(t)\big)}{1+\big(\partial_t f(t)\big) 
       \sigma\big(y_b(t)\big)}~.
\end{equation}
For the transformation (\ref{e:trafo}) to be regular, we have to require
$1+\dd_t s(t,y) =1 +\frac{df}{dt}(t)\sigma(y)\neq 0 ~ \forall (t,y)$.
If we choose $\sigma$ such that $ \partial_y \sigma(y_b(t))$ is bounded 
from below, $ 0<A<\partial_y \sigma\big(y_b(t)\big)$, this ansatz ensures 
the required asymptotics, $f(t)\ra 0$ for
$v(t)\ra 0$. In addition we need
\begin{equation}
\partial_y \sigma(y) |_{y = y_s} = 0~.
\end{equation}
The function $f(t)$ is determined by the differential equation
\begin{equation}\label{e:dif al}
\frac{df(t)}{dt}  +  \frac{1}{v(t)}\frac{\sigma'(y_b(t))}{\sigma(y_b(t))}  
 f(t)  +  \frac{1}{\sigma(y_b(t))} = 0~.
\end{equation}
A simple choice for $\sigma(y)$ is
\begin{equation}\label{e:sigma}
\sigma(y) = 1 + \frac{1}{\sigma_0}\left(1 - \frac{y}{y_s}\right)^2 ~,~~
  ~\sigma_0={\rm const.}, ~\sigma_0 > 1 ~,
\end{equation}
so that $ 1 \le \sigma(y) < 2$. With this, 
condition~(\ref{e:s condition 2}) is automatically satisfied. 
In addition, we want the brane collision, i.e. the bounce to happen at 
the fixed time $\tau=0$. For this we chose the initial condition
\begin{equation}
   f(t=0) = 0.
\end{equation}
Since $f(t)\ra 0$ for $v(t)\ra 0$,
the transformation (\ref{e:trafo}) satisfies 
\begin{equation}
\tau \ra t \;\;{\rm for }\;\;\;t\rightarrow \pm \infty\;\;\;{\rm and}\;\;\;
\tau(t=0,y) = 0.
\end{equation}
For the first of these equations we use that $v(t)\ra 0$ for 
$t\ra \pm\infty$ and the form of the differential equation~(\ref{e:dif al}).

\begin{figure}
\includegraphics[width=7cm]{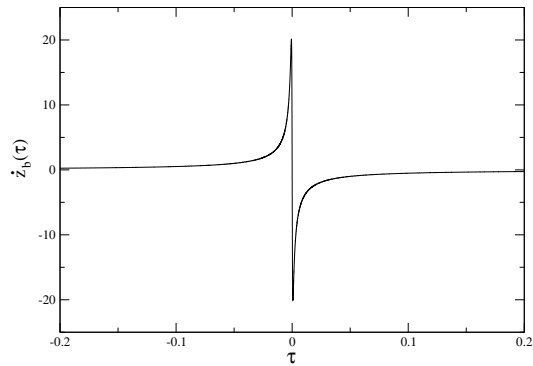}
\caption{\label{f:voft} The velocity in the new coordinates, 
$\frac{dz_b}{d\tau}$. Note that this is a coordinate velocity, not a 
physical quantity. It is easy to check that 
 $\tilde g_{AB}\frac{d\tilde x^A_b(\tau)}{d\tau}
\frac{d\tilde x^B_b(\tau)}{d\tau} <0$ at all times, hence the physical 
velocity remains timelike also if $\frac{dz_b}{d\tau}$ becomes larger 
than 1. The maximal velocity for this case is $v_{\max}=0.3$.}
\end{figure}

With (\ref{vtft}), $f(0)=0$ implies that the velocity vanishes at the brane, 
$v(0_-) =v_b =v(0_+) =0$. Hence the velocity does not jump from
a large value $v_b$ to $-v_b$ at the bounce but it evolves very rapidly but
smoothly from a high positive value $v_{\max}=v(-\epsilon)$ to a large 
negative value $-v_{\max}=v(\epsilon)$, $\epsilon > 0$ and small (see 
Fig.~\ref{f:voft}), like the regularized brane motion proposed in 
Section~\ref{s:setup}. Confirming that the results are independent of the 
choice of $\sigma_0$ is of course a crucial test.

The coordinate transformation maps the problem of a moving brane in a static 
bulk (\ref{e:bulk-metric}) onto the problem of a brane moving according to 
(\ref{e:zbrane motion}) in a time-dependent bulk. 
At first glance a further complication of the problem. 
Its benefits, however, will become clear in the next sections.  
The transformation of the metric is given explicitely in Appendix~A.

\subsection{Wave equation}
%
Transforming the Klein-Gordon operator (\ref{e:KG operator}) to the new 
coordinates $\tilde{x}^A$, we obtain the wave equation
\bea
\Big[ g_1(\tau,z) \partial_\tau^2 + g_2(\tau,z) \partial_\tau - 
2 s_2(\tau,z) \partial_z \partial_\tau  && \nonumber \\ \qquad + 
\frac{3}{z} \partial_z - \partial_z^2 + k^2 \Big] h (\tau,z) = 0~. &&
\label{e:new wave equation}
\eea
The definitions of the functions  $g_1(\tau,z)$,  $g_2(\tau,z)$ and 
$s_2(\tau,z)$ in terms of the coordinate 
transformation $s(t,y)$ are given in Appendix~\ref{A:coo}.  
These functions manifest that the bulk itself is now 
time-dependent and that the metric is no longer diagonal.
In Poincar\'e coordinates the non-triviality of the time-evolution of the 
perturbations is purely a consequence of the time-dependent junction condition, 
no time-dependent functions enter the wave equation (\ref{e:T-bulk-eq}). 
Our coordinate transformation which transforms the
generalized Neumann BC into a pure Neumann BC, 
induces explicite time-dependence in the wave equation 
itself. What is  important, however, is that in
(\ref{e:new wave equation}), in the instantaneous rest frame where we neglect 
time derivatives, we get the operator (\ref{e:mode}) of the original Bessel 
equation with normalized solutions (\ref{zero mode phi}) and 
(\ref{massive mode phi}). We just have to replace the variables $(t,y)$ by 
$(\tau,z)$.

Writing the action~(\ref{e:action h}) in terms of the 
new coordinates yields
\begin{eqnarray}
S &=& 4\int d\tau \frac{L^3}{2\kappa_5} \int d^3k \int_{z_b(\tau)}^{z_s} 
\frac{dz}{z^3} \frac{1}{1+ s_1}  \times  \label{e:action h new} \\
&&\hspace*{-1cm}
\Big[ g_1|\partial_\tau h|^2
-2\,s_2  {\rm Re}\left[ (\partial_\tau h)(\partial_z h^*)  \right]
\nonumber 
- |\partial_z h|^2 + k^2 |h|^2
 \Big]~.
\end{eqnarray}
Using the expressions for $s_1$, $s_2$, $g_1$ and $g_2$ given in 
Appendix~\ref{A:coo}, it is readily shown that the variation of 
(\ref{e:action h new}),  demanding Neumann boundary conditions
at the brane positions, leads to the wave equation (\ref{e:new wave equation}).

In the next subsections we take the action (\ref{e:action h new}) as the 
starting point to set up 
the dynamical Casimir effect formulation of graviton production
along the same lines as in \cite{DR,RD}. For the physical interpretation 
of gravitons we are using the fact that asymptotically, when 
the velocity of the brane goes to zero, the action (\ref{e:action h new}) 
and the wave equation (\ref{e:new wave equation}) reduce to (\ref{e:action h})
and (\ref{e:T-bulk-eq}), respectively. However, in the new coordinates, the 
junction conditions are always simple Neumann boundary conditions.
%
\subsection{Mode decomposition and Hamiltonian}
%
As a basis for a mode decomposition we chose the eigenfunctions 
$\{ \phi_\alpha(\tau,z)\}$ obtained by replacing $(t,y) \rightarrow (\tau,z)$
in (\ref{zero mode phi}) and (\ref{massive mode phi}).
As in \cite{RD} we call $\phi_0$ and $\phi_i$ the zero-mode and KK
mode solution, respectively. 
For a brane at rest, and hence $\tau = t$, the solutions 
$\phi_0$ and $\phi_i$ do indeed represent the physical zero mode and the 
KK modes, see, e.g. \cite{RS1}. When the brane is moving, however,
these solutions are 'instantaneous modes', provided that the boundary 
condition is Neumann. This approach is widely employed in the context of 
the dynamical Casimir effect, see 
\cite{Ruser:2004,Ruser:2005xg,Ruser:2006xg} and references therein.   
Here, working in the $(\tau,z)$-coordinates, the modes (\ref{zero mode phi}) 
and (\ref{massive mode phi}), are proper eigenfunctions respecting the full 
junction condition which we have reduced to a Neumann BC. 
At early and late times, i.e. asymptotically $|t| \rightarrow \infty$, where 
the brane velocity tends to zero, these eigenfunctions agree with the 
physical eigenfunctions corresponding to the zero mode and the KK modes.  
Since the eigenfunctions $\{\phi_\alpha(\tau,z)\}$ form a complete 
and orthonormal set.
and satisfy the correct junction conditions 
at both branes, we may decompose the graviton field in $\phi_\alpha$'s and
the pre-factors $q_{\alpha, {\bf k}} (\tau)$ become  canonical variables
which can then be quantized \cite{RD},
\begin{equation}
h(\tau,z, {\bf k}) = \sqrt{ \frac{\kappa_5}{L^3} }\sum_{\alpha=0}^{\infty} 
q_{\alpha, {\bf k}} (\tau) \phi_\alpha(\tau, z)~.
\label{e:mode decomposition}
\end{equation}
Our coordinate transformation and the expansion (\ref{e:mode decomposition})
satisfy two major requirements. First, the expansion 
(\ref{e:mode decomposition})
is consistent with the full junction condition (generalized Neumann BC). This 
overcomes the problem of our approach in \cite{DR,RD}. 
Secondly, even if at arbitrary times the $q_{\alpha, {\bf k}}$'s 
cannot a priori be identified with physical modes, asymptotically, 
i.e. when the brane moves very slowly, they do represent the independent 
physical graviton modes. This allows us to introduce
a proper notion of particles and vacuum states for asymptotic times. Initial 
and final vacuum states are then linked by the time-evolution 
of the  $q_{\alpha, {\bf k}}$'s  exactly as in \cite{RD}. 

We divide the wave equation~ (\ref{e:new wave equation}) by $g_1$ in order 
to isolate the second time derivative and insert the expansion 
(\ref{e:mode decomposition}). Note that $g_1\ra 1$ for $|t|\ra\infty$ and
for a sufficiently large choice of $\sigma_0$, $g_1>0$ at all times. As we 
shall see below, this is also needed for the Hamiltonian to be positive at 
all times. Inserting the expansion (\ref{e:mode decomposition}) into 
(\ref{e:new wave equation}), multiplying it by $\phi_\beta$ and integrating 
over $2\,\int_{z_b(\tau)}^{z_s} dz/z^3$ leads to a system of differential 
equations for the $q_{\alpha, {\bf k}}$
which has the same form as the one of Refs.~\cite{DR,RD},
\begin{equation}
\ddot{q}_{\alpha, {\bf k}}(\tau) + \sum_\beta 
\left[ A_{\beta\alpha} (\tau)\dot{q}_{\beta, {\bf k}} 
(\tau)+ B_{\beta\alpha}(\tau) q_{\beta, {\bf k}}(\tau) \right] = 0~.
\label{e:deq for q from wave equation}
\end{equation}
The explicite expressions for the time-dependent coupling matrices 
$ A_{\beta\alpha}$ and $ B_{\beta\alpha}$ are given by integrals over 
the bulk which are rather 
cumbersome. The details can be found in Appendix~\ref{A:pert}.
Inserting the expansion (\ref{e:mode decomposition}) into the action 
(\ref{e:action h new}) we obtain the Lagrangian $L(\tau)$ in terms of the 
variables $ q_{\alpha, {\bf k}}(\tau)$. We can then define the canonical momenta
$p_{\alpha, {\bf k}} = \partial L/\partial\dot{q}_{\alpha,{\bf k}}$
from which, by means of a Legendre transformation, we derive the Hamiltonian
\begin{eqnarray}
 H(\tau) = \frac{1}{2}\int d^3k 
\sum_{\alpha\beta} \Big[
 p_{\alpha, {\bf k}} \,E^{-1}_{\alpha\beta} \,p_{\beta, {\bf -k}}
 \hspace*{1.5cm}  && \nonumber \\
+ q_{\alpha, {\bf k}} \left[ \frac{1}{2} 
\left( \omega_{\alpha, k}^2 (\tau) + \omega_{\beta, k}^2(\tau) \right ) 
\delta_{\alpha\beta}
+ V_{\alpha\beta} \right] q_{\beta, {\bf -k}} && \nonumber \\
- (M_{\beta\alpha} - S_{\beta\alpha})\left[ q_{\beta, {\bf k}} 
p_{\alpha, {\bf k}}   
+p_{\alpha, {\bf k}}   q_{\beta, {\bf k}}
\right]  \Big]\,. \hspace*{0.5cm} &&
\label{e:H}
 \end{eqnarray}
The matrices $E^{-1}_{\al\beta}$, $V_{\alpha\beta}$, $M_{\beta\alpha}$
and  $ S_{\beta\alpha}$ are given explicitely in Appendix~\ref{A:pert}.
It is important to note that $E^{-1}_{\al\beta}$ is positive definite as long 
as $g_1>0$ and $1+s_1>0$, which we have to  require for our approach to be
consistent. In the old treatment, $E_{\al\beta}$ was the identity matrix and
the couplings $S_{\beta\alpha}$ and  $V_{\alpha\beta}$ were missing.  They are
due to the time-dependence of the bulk spacetime in the new coordinates and 
therefore originate from the term $v\dd_t$ of the boundary condition in 
Poincar\'e coordinates. The coupling matrix $M_{\beta\alpha}$ which is also 
present in our previous treatment comes from the time dependent Neumann BC.
Finally, the time dependence of the bulk volume $z_s-z_b(\tau)$, induces
the time dependence in the frequency $\om_{\al,k}$ (squeezing effect, 
see \cite{RD}).

All the coupling matrices tend to zero when $v\ra0$.  But we have not been 
able to show that the new couplings, $E^{-1}_{\al\beta} 
-\de_{\al\beta}$, $V_{\alpha\beta}$ and $S_{\beta\alpha}$ are parametrically
smaller than  $M_{\beta\alpha}$, e.g. that they are of order $v^2$. Therefore, 
the result that the particle production obtained in our previous treatment~\cite{DR,RD} 
is not modified if the velocity is
sufficiently low is not evident and has to be checked numerically. 
\\

%
\subsection{Quantum Generation of Gravitons \label{s:QM}}
%
The quantization procedure goes along the same lines as in \cite{RD}.
The canonical variables $q_{\alpha,{\bf k}}(\tau),~ p_{\alpha,{\bf k}}(\tau)$
and the Hamiltonian $H(\tau)$ are promoted to operators
$\hat{q}_{\alpha,{\bf k}(\tau)},~ \hat{p}_{\alpha,{\bf k}}(\tau)$
and $\hat{H}(\tau)$, subject to the usual commutation relations. 
In the Heisenberg picture where the time evolution of an operator $\hat{O}$ 
is determined by $$\dot{\hat{O}}(\tau) = i [\hat{H}(\tau),\hat{O}(\tau)] + 
\left( \frac{\partial \hat{O}(\tau)}{\partial \tau} \right)_{\rm expl.}\,,$$
the operators $\hat{q}_{\alpha,{\bf k}}(\tau)$ 
and $\hat{p}_{\alpha,{\bf k}}(\tau)$
satisfy the same Hamiltonian equations of motion as their classical 
counterparts, i.e.
\begin{equation}
\dot{q}_{\alpha, {\bf k}} = \frac{\partial H}{\partial p_{\alpha, {\bf k}}}\;,\;\;
\quad
\dot{p}_{\alpha, {\bf k}} = - \frac{\partial H}{\partial q_{\alpha, {\bf k}}}
\label{e:Hamilton eq 1}
\end{equation}
Remember that we assume that asymptotically, $|t| \rightarrow \infty$, the 
brane is at rest, i.e. the brane velocity vanishes and both coordinate systems 
agree, $\tau = t$. We extend this notion of asymptotic behavior by 
introducing two times, $t_{\rm in}$ and
$t_{\rm out}$, and we shall assume that the brane is at rest for 
$t \le t_{\rm in}$ and $t \ge t_{\rm out}$, respectively.
This corresponds to a scenario where the motion of the brane is switched 
on and off at finite times.
Such a brane dynamics may seem rather artificial from a physical point of view,
but what is important for us is that before $t_{\rm in}$ and after 
$t_{\rm out}$ no significant particle creation takes place. Numerically, we 
test this by varying $ t_{\rm in}$ and $t_{\rm out}$ and choosing them large 
enough so that the particle number is independent of the value chosen.

In the $(\tau,z)$-coordinates,  the brane is then at rest for times 
$\tau \equiv t \le \tau_{\rm in} \equiv t_{\rm in}$ and 
$ \tau \equiv t \ge \tau_{\rm out}\equiv t_{\rm out}$, respectively.
When the brane velocity is zero, the matrix 
$E_{\alpha\beta}(\tau)$ defined in Appendix~\ref{A:pert}
becomes the identity, $E_{\alpha\beta}(\tau)~  
{\rightarrow}_{|\tau| \rightarrow \infty} ~ \delta_{\alpha\beta}$, and all other 
matrices which represent the coupling 
terms vanish identically in this limit.
Consequently, for asymptotic times the Hamiltonian reduces to the 
familiar form of a collection of independent harmonic oscillators,
\begin{equation}
\hat{H}^{\rm in / out} = \frac{1}{2}\int d^3k \sum_{\alpha} \left[
|\hat{p}_{\alpha, {\bf k}}|^2 + \left( \omega_{\alpha,k}^{\rm in/out} \right)^2 
|\hat{q}_{\alpha, {\bf k}}|^2 \right]
\label{e:asymp H}
\end{equation}
with 
\begin{equation}
\hat{p}_{\alpha, {\bf k}} = \dot{\hat{q}}_{\alpha, {\bf -k}} \, .
\end{equation}
 We have introduced the notation
\begin{equation}
\omega_{\alpha, k}^{\rm in} \equiv \omega_{\alpha, k} (\tau \le \tau_{\rm in})\;,\;\;\
\omega_{\alpha, k}^{\rm out} \equiv \omega_{\alpha, k} (\tau \ge \tau_{\rm out})~.
\end{equation}
Following \cite{RD}, we  decompose $\hat{q}_{\alpha, {\bf k}}$ in creation and 
annihilation operators,
\begin{equation}
\hat{q}_{\alpha, {\bf k}} (\tau) = 
\sum_\beta \frac{1}{\sqrt{2\,\omega_{\beta, k}^{\rm in}}}
\left[ \hat{a}^{\rm in}_{\beta, {\bf k}} \epsilon^{(\beta)}_{\alpha, k} (\tau) + 
\hat{a}^{{\rm in}\dagger}_{\beta, -{\bf k}} 
\epsilon^{{(\beta)}*}_{\alpha, k} (\tau) \right]~,
\end{equation}
which are defined via $\hat{a}^{\rm in}_{\alpha,{\bf k}} |0, {\rm in}\rangle 
= 0 \;\;\forall \;\;\alpha, {\bf k}$.
The initial vacuum state $|0, {\rm in}\rangle$ is the ground state of the 
Hamiltonian (\ref{e:asymp H}) for times $\tau \le \tau_{\rm in}$.
This initial state is linked to the final vacuum state defined by  
$\hat{a}^{\rm out}_{\alpha,{\bf k}} |0, {\rm out}\rangle = 0 \,;\;\forall \;\;
\alpha, {\bf k}$, by means of a Bogoliubov transformation (see \cite{RD})
\begin{equation}
\hat{a}^{\rm out}_{\beta,{\bf k}} = \sum_\alpha \left[{\cal A}_{\alpha\beta,k} (\tau_{\rm out}) 
\hat{a}^{\rm in}_{\alpha,{\bf k}} + {\cal B}^*_{\alpha\beta,k} (\tau_{\rm out}) 
\hat{a}^{{\rm in}\dagger}_{\alpha,{\bf -k}}\right]
\end{equation}
which determines the number of produced gravitons  (for each polarization)
\begin{equation}
N_{\al, k}^{\rm out} = \sum_\beta |{\cal B}_{\beta\alpha,k}(\tau_{\rm out})|^2~.
\label{e:final number}
\end{equation}
As we have discussed in detail in \cite{RD}, the graviton number after the time 
$\tau_{\rm out}$,  (\ref{e:final number}), represents a physically meaningful 
quantity . 
%
\section{Numerical Results}\label{s:num}

\begin{figure}[ht]
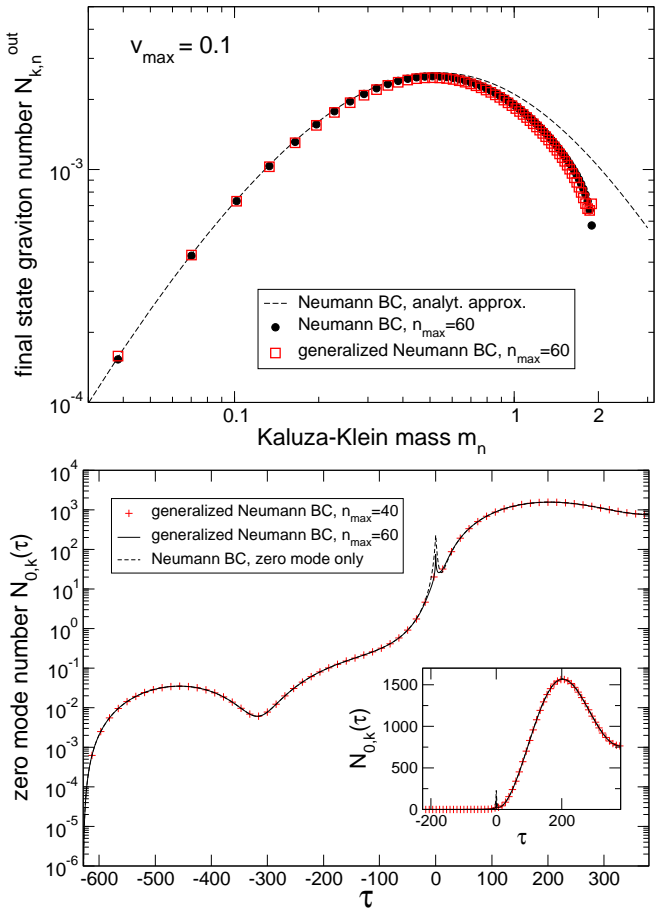

\begin{center}
\includegraphics[height=6cm]{KK_k001ys100vb01_new.eps}
\\
\includegraphics[height=6cm]{KK_k001ys100vb01_zm_new.eps}
\caption{The final graviton spectrum for three-momentum $k=0.01$,
brane separation $y_s=100$ and bounce velocity $v_{\max} = 0.1$. 
The top panel shows the final KK mode spectrum and the lower panel 
depicts the time evolution of the zero mode.  What we plot here is a 
kind of instantaneous particle number (see Appendix C of \cite{RD}).
The numerical result for the KK spectrum is compared with the old one 
(shown in Fig.~13 of \cite{RD}). 
Like there, lengths are in units of $L$ and momenta/masses in the units 
of $L^{-1}$. For low velocities $v_{\max}\le 0.1$ the new spectra
(generalized Neumann BC) are identical with the old ones (Neumann BC) within 
the numerical error which are estimated by the Bogoliubov test (see Appendix).
$N_{0,k}(\tau)$ is shown for two cut-off parameters $n_{\rm max}$ to 
underline stability of the solution. \label{figure 1}}
\end{center}
\end{figure}

\begin{figure}[ht]
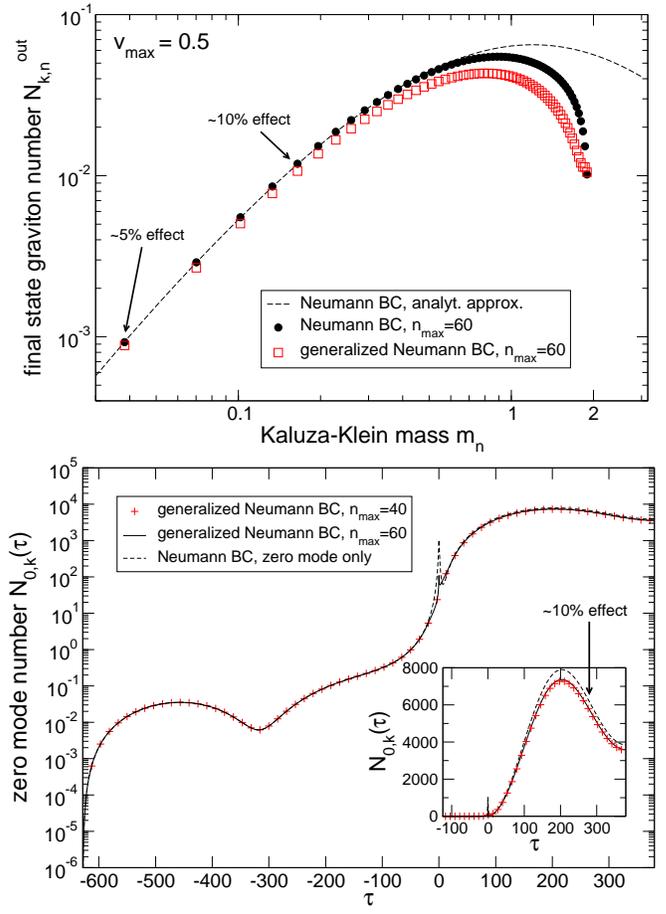

\begin{center}
\includegraphics[height=6cm]{KK_k001ys100vb05_new.eps}
\\
\includegraphics[height=6cm]{KK_k001ys100vb05_zm_new.eps}
\caption{As Figure~\ref{figure 1} but for $v_{\max}=0.5$. For this
velocity we do see a difference of about 10\% between the 
previous, inconsistent approach and the new generalized Neumann BC for both, 
the KK modes as well as the zero mode. Again, the 4d graviton number is 
shown for two cut-off parameters $n_{\max}$ to indicate numerical stability. .
\label{figure 1b}}
\end{center}
\end{figure}

\begin{figure}[ht]
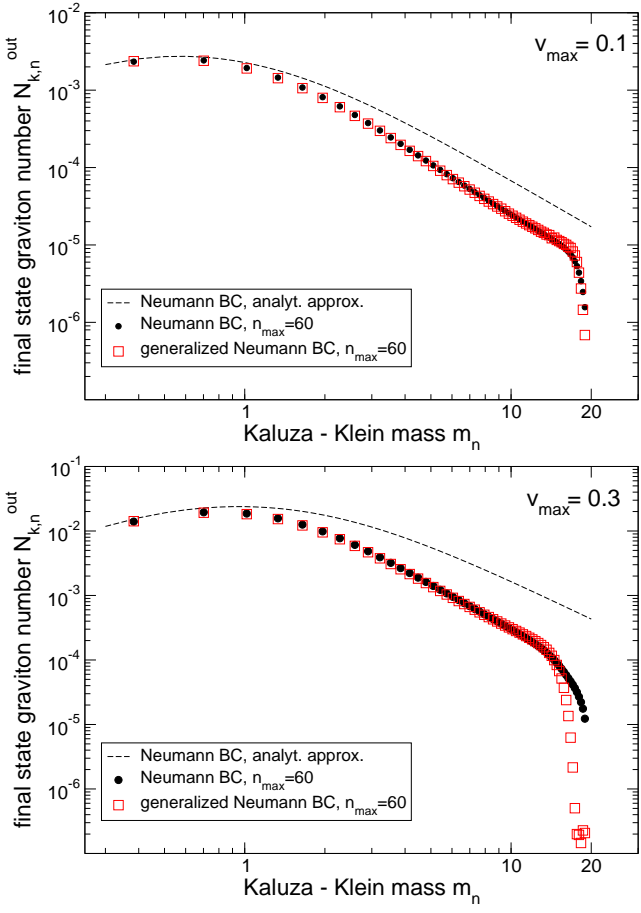

\begin{center}
\includegraphics[height=6cm]{KK_k01ys10vb01_new.eps}
\\
\includegraphics[height=6cm]{KK_k01ys10vb03_new.eps}
\caption{Final graviton spectra for three-momentum $k=0.1$ and
brane separation $y_s=10$ for the bounce velocities 
$v_{\max} = 0.1$ and $0.3$. Again, the new (generalized Neumann BC) numerical 
results are compared with the ones of the previous inconsistent approach 
(Neumann BC),  see Fig.~15 of \cite{RD},
and the agreement is excellent in the regime $m_n<13$,
where the numerics can be trusted.
\label{f:more}}
\end{center}
\end{figure}

In order to solve the equations of motion 
(\ref{e:deq for q from wave equation}) numerically, we transform the system 
to a first order system and introduce a mass cutoff, $n_{\max}$, i.e. we 
neglect all modes with masses higher than $m_{n_{\max}}$, in other words
$q_\al=0$ for $\al>n_{\max}$, along the lines explained in detail in 
Ref.~\cite{RD}. Modes close to this cutoff are of course seriously affected 
by it as is seen in Figs.~\ref{figure 1}, \ref{figure 1b} and \ref{f:more}.
We have tested the stability of the results for modes $n\ll n_{\max}$ by varying
the cutoff. The stability of the zero mode is illustrated in the  lower 
panel of Figs.~\ref{figure 1} and~\ref{figure 1b}.
Typically modes with $n\lsim 0.7n_{\max}$ can be trusted. An indication for 
this is also the Bogoliubov test shown in Fig.~\ref{figure ap1}
and discussed in Ref.~\cite{RD}.

The first order system is given explicitly in 
Appendix~\ref{a:bogo} and differs from the original one in 
\cite{RD} only by additional mode couplings.
We have compared our new results 
with those of Ref.~\cite{DR,RD} and find excellent agreement at low 
velocity, $v_{\max}\lsim 0.1$. This is illustrated in Figs.~\ref{figure 1} 
and~\ref{figure 1b}. At bounce velocities $v_{\max}\gsim 0.5$ we do find 
differences
as expected, but these results cannot be taken literally since for these
velocities the low energy evolution of the scale factor adopted in this work 
is no longer sufficient. We will present the full high velocity results in
a forthcoming paper~\cite{pap2}.


The agreement between the old results for pure Neumann boundary 
conditions and the new ones with the generalized Neumann boundary 
conditions is similar for other values of $y_s$ and $k$.

In Fig.~\ref{f:more} we show the KK spectra for $v_{\max}=0.1$ and $v_{\max}=0.3$
for the wave number $k=0.1$ and position of the static brane, $y_s=10$.
In this case, the analytic approximation derived in Ref.~\cite{RD} which
is valid for $m_n<1$ can only be trusted for the first two modes. 
The slight difference between the old and the new spectra towards the end, 
i.e. for  $m_n > 10$, is due to changes how we numerically evolve the 
solutions through the bounce (see Appendix C). This affects the 
sensitivity of the solutions to the 
cut-off. What we observe here as a slight bending of the spectrum for 
generalized Neumann BC's is also found in our previous approach if we 
increase the number of modes; compare to the $n_{\rm max}=100$ results shown 
in Fig.~15 of~\cite{RD} and the discussion related to Fig.~25 of \cite{RD}. 
The drop in the final part of the spectrum is just an artefact 
of the finite cut-off (see~\cite{RD} for a detailed discussion).

\section{Conclusions}\label{s:con}
In this paper we have derived a method to calculate graviton production 
in bouncing AdS$_5$ braneworlds
by the dynamical Casimir effect taking into account the full generalized 
Neumann boundary condition. We have achieved this by transforming to a 
new time coordinate, in which the generalized Neumann BC become 
ordinary Neumann BC. We have shown numerically that for low bounce
velocities, $v_{\max}\lsim 0.1$, the number of generated particles agrees with 
the one from the simpler treatment which neglects the velocity term in 
the boundary condition.
Since this term is of first order in the velocity, we believe that our
result is not obvious. Furthermore, the method developed in this work can
be used to calculate particle creation for branes moving at arbitrarily high
velocities. In this case, one will have to 
take into account the  modification of the Friedmann equation at high 
energy, $HL\gsim 1$. This is the goal of a forthcoming paper~\cite{pap2}.

In this work we have not derived new physical results, but we have 
developed a new, fully consistent method to calculate graviton
production due to the motion of a braneworld. Our method overcomes a 
shortcoming of our previous treatment~\cite{DR,RD}, and we have verified that 
at low brane velocity, $v_{\max}\lsim 0.3$ the previous results are not affected.

\section*{Acknowledgment}
This work is supported by the Swiss National Science Foundation.
The numerical simulations have been carried out on the Myrinet cluster 
of Geneva University. RD thanks the Galileo Galilei Institute of  theoretical 
physics, where this work was finalized, for hospitality. 

\begin{appendix}

\section{The coordinate transformation}\label{A:coo}
The Jacobian ${\bf T}$ of the transformation 
$$(t,y) \mapsto (\tau=t+s(t,y),z=y)$$ reads
\be\label{jacob}
{\bf T} = \frac{\partial (\tau,{\bf x},z)}{\partial(t,{\bf x},y)} = \left(
\begin{array}{cccccc}
1 + \partial_t s & 0 & 0 & 0& \partial_y s\\
0 & 1 & 0 & 0 & 0\\
0 & 0 & 1 & 0 & 0\\
0 & 0 & 0 & 1 & 0\\
0 & 0 & 0 & 0 & 1\\ 
\end{array}
\right )\,,
\ee
and its inverse is 
\[
{\bf T}^{-1} = \frac{\partial (t,{\bf x},y)}{\partial(\tau,{\bf x},z)}
= \left(
\begin{array}{cccccc}
\frac{1}{1+ \partial_t s} & 0 & 0 & 0&\frac{ -\partial_y s}{1+ \partial_t s}\\
0 & 1 & 0 & 0 & 0\\
0 & 0 & 1 & 0 & 0\\
0 & 0 & 0 & 1 & 0\\
0 & 0 & 0 & 0 & 1\\ 
\end{array} \right)~.
\]
Under this coordinate transformation the AdS$_5$ metric in Poincar\'e 
coordinates given in~(\ref{e:bulk-metric}) transforms to 
\begin{eqnarray}
\tilde{g}_{AB}d\tilde{x}^A d\tilde{x}^B &=& \left( ({\bf T}^{-1})^T g {\bf T}^{-1}\right)_{AB}d\tilde{x}^A d\tilde{x}^B \nonumber \\
&=& 
\frac{L^2}{z^2}\Big[  \frac{1}{(1 + s_1(\tau,z))^2}
\Big(-d\tau^2 \\ && \hspace*{-1.2cm} +2 s_2(\tau,z)d\tau dz  + g_1(\tau,z)dz^2\Big) 
+\de_{ij}dx^idx^j\Big]~.\nonumber
\label{e:new metric expl}
\end{eqnarray}
We introduce the functions
\begin{eqnarray}
s_1(\tau,z) &=& (\partial_t s)\big(t(\tau,z),z\big)  
\label{e:s1}\\
s_2(\tau,z) &= &\left. (\partial_z s)\right|_{t=\mr{const}}\big(t(\tau,z),z\big) 
\label{e:s2}\\
s_{11}(\tau,z) &=& (\partial_t^2 s) \big(t(\tau,z),z\big)
\label{e:s11}\\
s_{22}(\tau,z) &=&\left. (\partial_z^2 s)\right|_{t=\mr{const}} \big(t(\tau,z),z\big) 
\label{e:s22}
\end{eqnarray}
and
\begin{eqnarray}
g_1(\tau,z) &=& \big(1 + s_1(\tau,z)\big)^2 - s_2(\tau,z)^2 
\label{e:g1} \\
g_2(\tau,z) &=& s_{11}(\tau,z) - s_{22}(\tau,z) + \frac{3}{z} s_2(\tau,z) ~,
\label{e:g2}
\end{eqnarray}
of which $g_1$ and $g_2$ will be used in Appendix~\ref{A:pert}. 
\\
The determinant is 
\begin{equation}
\tilde{g} = {\rm det} (\tilde{g}_{AB}) = - \left(\frac{L}{z}\right)^{10} \frac{1}{(1+s_1(\tau,z))^2}~.
\end{equation}

\section{Details on evolution equations}\label{A:pert}
\subsection{Wave equation}
The coupling matrices which determine the mode evolution 
equation~(\ref{e:deq for q from wave equation}) are given in
terms of the following bulk integrals:

\begin{eqnarray}
A_{\alpha\beta}(\tau)  &=& 2\,\int_{z_b(\tau)}^{z_s} \frac{dz}{z^3} 
\left[ 2 \dot{\phi}_\alpha + \frac{g_2(\tau,z)}{g_1(\tau,z)}\phi_\alpha(\tau,z) 
\right. \nonumber \\ &&  \left. \qquad
- \frac{2 s_2(\tau,z)}{g_1(\tau,z)} \phi_\alpha^\prime (\tau,z)\right] 
\phi_\beta (\tau,z)\\
B_{\alpha\beta} (\tau) &=& 2\,\int_{z_b(\tau)}^{z_s} \frac{dz}{z^3} \left[  
\ddot{\phi}_\alpha(\tau,z) + \frac{g_2(\tau,z)}{g_1(\tau,z)} 
\dot{\phi}_\alpha(\tau,z)
\right. \nonumber \\ &&  \left. \hspace{-1.8cm}
 - \frac{2 s_2(\tau,z)}{g_1(\tau,z)} \dot{\phi}_\alpha^\prime (\tau,z) + 
\frac{\omega_{\alpha,k}^2(\tau)}{g_1(\tau,z)} \phi_\alpha(\tau,z) \right] 
\phi_\beta (\tau,z)~
\end{eqnarray}
with
\begin{equation}
\omega_{\alpha, k} (\tau) = \sqrt{m_\alpha^2(\tau) + k^2}~.
\end{equation}
The over-dot denotes the derivative w.r.t. the time $\tau$ and a prime stands 
for the derivative w.r.t. the coordinate $z$.
Compared to our former work \cite{RD}, the present problem
is more complicated due to the additional couplings which are
caused by the time-dependence of the bulk spacetime. 
Also the Lagrangian and Hamiltonian
equations for $q_{\al,k}$ are more complicated. Furthermore, the functions 
$s_1,s_2,g_1$ and $g_2$ are only known numerically. This induces additional
numerical difficulties. Note also that it is important that $g_1$ does not pass 
through zero for these integrals to be well defined, hence 
$g_1(\tau,z)>0~\forall \tau,z$. This is, however, easily achieved 
with our ansatz~(\ref{e:seqfsi}, \ref{e:sigma}) for $s(t,y)$ if we choose 
$\sigma_0$ sufficiently large.
%
\subsection{Lagrangian and Hamiltonian formulation}
%
Inserting the expansion (\ref{e:mode decomposition}) into the action 
(\ref{e:action h new}) leads (for each of the polarizations) to the Lagrangian 
\begin{eqnarray}
L(\tau) &=& \frac{1}{2} \int d^3k \sum_{\alpha\beta} \Big[ 
E_{\alpha\beta} \dot{q}_{\alpha,{\bf k}}\dot{q}_{\beta,{\bf -k}} 
 \nonumber \\  && 
+ ({\cal M}_{\alpha\beta}  - {\cal K}_{\alpha\beta} )
(q_{\alpha,{\bf k}}\dot{q}_{\beta,{\bf -k}}  + 
q_{\alpha,{\bf -k}}\dot{q}_{\beta,{\bf k}} )  \nonumber \\
 && \nonumber + ({\cal N}_{\alpha\beta} - {\cal P}_{\alpha\beta} - 
Q_{\alpha\beta} -\omega^2_{\alpha\beta,k} )
q_{\alpha, {\bf k}} q_{\beta, -{\bf k}} 
\Big]~
\label{e:Lagrangian new}
\end{eqnarray}
containing several time-dependent coupling terms. In detail, these read
\begin{eqnarray*}
E_{\alpha\beta}(\tau) &=& 2 \int_{z_b(\tau)}^{z_s} \frac{dz}{z^3} \frac{g_1(\tau,z)}{1+s_1(\tau,z)}
\phi_\alpha(\tau,z) \phi_\beta(\tau,z)
\\
{\cal M}_{\alpha\beta}(\tau) &=& 2 \int_{z_b(\tau)}^{z_s} \frac{dz}{z^3} 
\frac{g_1(\tau,z)}{1+s_1(\tau,z)}
\dot{\phi}_\alpha(\tau,z) \phi_\beta(\tau,z)
\\
{\cal N}_{\alpha\beta}(\tau) &=& 2 \int_{z_b(\tau)}^{z_s} \frac{dz}{z^3} 
\frac{g_1(\tau,z)}{1+s_1(\tau,z)}
\dot{\phi}_\alpha(\tau,z) \dot{\phi}_\beta(\tau,z)
\\
{\cal K}_{\alpha\beta}(\tau) &=& 2 \int_{z_b(\tau)}^{z_s} \frac{dz}{z^3} 
\frac{s_2(\tau,z)}{1+s_1(\tau,z)}
\phi^\prime_\alpha(\tau,z) \phi_\beta(\tau,z)
\\
{\cal P}_{\alpha\beta}(\tau) &=& 2 \int_{z_b(\tau)}^{z_s} \frac{dz}{z^3} 
\frac{s_2(\tau,z)}{1+s_1(\tau,z)}
\big[ \dot{\phi}_\alpha(\tau,z) \phi_\beta^\prime (\tau,z) \\ && + 
 \phi^\prime_\alpha(\tau,z) \dot{\phi}_\beta (\tau,z) \big]
\\
{\cal Q}_{\alpha\beta}(\tau) &=&  \int_{z_b(\tau)}^{z_s} \frac{dz}{z^3} 
\frac{s_1^\prime(\tau,z)}{(1+s_1(\tau,z))^2} 
\big[ \phi_\alpha(\tau,z) \phi_\beta^\prime (\tau,z) \\ && + 
 \phi^\prime_\alpha(\tau,z) \phi_\beta (\tau,z) \big]
\\
\omega^2_{\alpha\beta, k} (\tau) &=& 2 \left[ \frac{1}{2} 
\left( m_\alpha^2(\tau) + m_\beta^2(\tau) \right)  + k^2\right]\times \\ &&
 \int_{z_b(\tau)}^{z_s} \frac{dz}{z^3}\frac{\phi_\alpha(\tau,z) 
\phi_\beta(\tau,z)}{1+s_1(\tau,z)}~.
\end{eqnarray*}
 Since we require $g_1(\tau,z)>0$ and $1+s_1(\tau,z)>0$,
the matrix $E_{\al\beta}$ is positive definite. This is important for the above
Lagrangian to lead to consistent second order equations of motion for the
variables $q_{\al,{\bf k}}$ (no ghosts).

The equation of motion for the canonical variables obtained from the 
Euler--Lagrange equations become
\begin{eqnarray}
\sum_{\alpha} \Big[ E_{\alpha\gamma}\ddot{q}_{\alpha, {\bf k}} +  
\dot{E}_{\alpha\gamma}\dot{q}_{\alpha, {\bf k}} 
 + \left( \dot{{\cal M}} - \dot{{\cal K}}  \right)_{\alpha\gamma} 
q_{\alpha, {\bf k}} && \nonumber \\  
+ \left[\left( {\cal M} - {\cal K}  \right)_{\alpha\gamma} -  
\left( {\cal M} - {\cal K}  \right)_{\gamma\alpha}\right] 
\dot{q}_{\alpha, {\bf k}}  
&& \nonumber \\
-\left( {\cal N}_{\alpha\gamma} - {\cal P}_{\alpha\gamma} - Q_{\alpha\gamma} -
\omega_{\alpha\gamma, k}  \right) q_{\alpha, {\bf k}} 
\Big] = 0~. && 
\label{e:deq for q from L}
\end{eqnarray}
Note that all the matrices introduced above apart from $E_{\al\beta}$ 
 tend to zero when $v\ra 0$, i.e. for $|\tau|\ra \infty$.
In this limit $E_{\al\beta}$ tends to $\de_{\al\beta}$  so that
Eq.~(\ref{e:deq for q from L}) becomes the free, uncoupled mode evolution 
equation in this limit as is expected.
Introducing the canonically conjugate variables 
\begin{equation}
p_{\alpha, {\bf k}} = \frac{\partial L}{\partial \dot{q}_{\alpha, {\bf k}}}
=
\sum_\beta \left[ E_{\alpha\beta} \dot{q}_{\beta, {\bf -k}} + 
({\cal M}_{\beta\alpha}  - {\cal K}_{\beta\alpha}) q_{\beta, {\bf -k}}\right]
\end{equation}
leads by means of a Legendre transformation to the Hamiltonian (\ref{e:H})
with coupling matrices
\begin{eqnarray}
V_{\alpha\beta} (\tau) &=& 2 \int_{z_b(\tau)}^{z_s} \frac{dz}{z^3} \frac{1}{1+s_1}
\Big[ \left( \frac{s_2^2}{g_1} - s_1\right) 
\phi^{'}_{\alpha}(\tau,z) \phi^{'}_{\beta}(\tau,z) \nonumber \\ &&
\qquad -k^2 s_1(\tau,z) \phi_\alpha(\tau,z) \phi_\beta(\tau,z)\Big]
\\
M_{\alpha \beta}(\tau) &=& 2 \int_{z_b(\tau)}^{z_s} \frac{dz}{z^3} 
\dot{\phi}_\alpha(\tau,z) \phi_\beta(\tau,z) 
\\
S_{\alpha \beta} (\tau) &=& 2 \int_{z_b(\tau)}^{z_s} \frac{dz}{z^3} 
\frac{s_2}{g_1}\phi^\prime_\alpha(\tau,z) \phi_\beta(\tau,z) ~.
\end{eqnarray}
Thereby $E_{\alpha\beta}^{-1}$ is the inverse of  $E_{\alpha\beta}$, i.e.
\begin{equation}
E_{\alpha\beta}^{-1} = 2 \int_{z_b(\tau)}^{z_s} \frac{dz}{z^3} \frac{1+s_1(\tau,z)}{g_1(\tau,z)}
\phi_\alpha (\tau,z) \phi_\beta(\tau,z)~.
\end{equation}
The Hamilton equations
\begin{equation}
\dot{q}_{\alpha, {\bf k}} = \frac{\partial H}{\partial p_{\alpha, {\bf k}}}\;,\;\;
\dot{p}_{\alpha, {\bf k}} = - \frac{\partial H}{\partial q_{\alpha, {\bf k}}}
\label{eap:Hamilton eq 1}
\end{equation}
then provide the equations of motion  
for the variables $q_{\alpha, {\bf k}}$ and $p_{\alpha, {\bf k}}$.

Using certain relations of the coupling matrices following from the 
completeness (\ref{e:complete}) and ortho-normality (\ref{e:orthonormal}) 
of the functions $\phi_\alpha$ and the properties of the functions 
$s_1, s_2,  s_{11}$ and $s_{22}$ one can show that the 
three systems of equations
(\ref{e:deq for q from wave equation}), (\ref{e:deq for q from L}) 
and the Hamilton equations (\ref{eap:Hamilton eq 1}) are 
consistent with each other, i.e. one system follows from the other one.  
This seems to be at first sight a rather trivial statement but we have to 
remind the reader that this is not the case in our previous work \cite{DR,RD} 
as we have discussed in detail in Section II. D of Ref.~\cite{RD}. 
The new coupling matrices $V_{\alpha\beta}$,  $S_{\alpha\beta}$ and  
$E_{\alpha\beta} -\de_{\alpha\beta}$ are missing in our previous work. 
Even though they do become very small when the brane velocity becomes small,
it is not evident that these new terms must be smaller than
e.g. $M_{\alpha\beta}$, which also tends to zero with $v$. In other words,
it is not straight forward to show that these contributions are, e.g., of
order $v^2$.
\subsection{Bogoliubov coefficients} \label{a:bogo}

Performing the quantization as in Ref.~\cite{RD} we again transform to a 
first order system of equation. In the new coordinates  the system of 
equations (3.34), (3.35) of \cite{RD} is replaced by 
\begin{eqnarray} \label{e:zetadot}
\dot{\xi}_\alpha^{(\gamma)} (\tau)&=& \sum_\beta \Big\{ - 
\left[ ia_{\alpha\beta}^+(\tau) + c_{\alpha\beta}^-(\tau)\right]
\xi_\beta^{(\gamma)}(\tau) \nonumber \\ && + \left[ ia_{\alpha\beta}^-(\tau) -
 c_{\alpha\beta}^+(\tau)\right] \eta_\beta^{(\gamma)}(\tau)
\Big\} \\   \label{e:etadot}
\dot{\eta}_\alpha^{(\gamma)} (\tau) &=& \sum_\beta \Big\{ 
\left[ ia_{\alpha\beta}^+ (\tau) -  c_{\alpha\beta}^-(\tau)\right]
\eta_\beta^{(\gamma)} (\tau) \nonumber \\ &&  - \left[ ia_{\alpha\beta}^-(\tau) + 
c_{\alpha\beta}^+(\tau)\right] \xi_\beta^{(\gamma)}(\tau) \Big\}
\end{eqnarray}
where
\bea
a^\pm_{\alpha\beta}(\tau) &=& \frac{1}{2} \Big[ \omega^{\rm in}_{\beta, k} 
E^{-1}_{\alpha\beta}(\tau) \pm \frac{1}{\omega^{\rm in}_{\alpha, k}} \Big( 
\frac{1}{2} (\omega_{\alpha, k}^2(\tau) \nonumber \\  &&  \quad + 
\omega_{\beta, k}^2(\tau))\delta_{\alpha \beta} + V_{\alpha\beta}(\tau) \Big) 
\Big] 
\eea
\bea
c_{\alpha\beta}^\pm (\tau) &=& \frac{1}{2} \Big[ M_{\beta\alpha}(\tau) - 
S_{\beta\alpha} (\tau) \nonumber \\  && \pm 
\frac{\omega_{\beta, k}^{\rm in} }{\omega_{\alpha, k}^{\rm in} } \left(
M_{\alpha\beta}(\tau) - S_{\alpha\beta} (\tau)\right)  \Big]~.
\eea
(Note that in Ref.~\cite{RD} a factor of two is missing in the expression for 
$M_{ij}^{\rm N}$  in Eq. (B8), a simple misprint.)  

\section{Numerics}
To compute the graviton spectra we have adapted the code described
in Ref.~\cite{RD} to the new problem. Apart from calculating the new coupling 
matrices we also have to solve the differential equation~(\ref{e:dif al}) 
numerically to calculate the coordinate transformation and its inverse in 
order to determine $z_b(\tau)$ via the implicit 
equation~(\ref{e:zbrane motion}) as well as the functions $s_1(\tau,z)$, 
$s_2(\tau,z)$, $g_1(\tau,z)$ and $g_2(\tau,z)$ which enter the integrals for 
the coupling matrices. For numerical purposes we have smoothed the function 
$y_b(t)$. Due to this implicit nature of the coordinate transformation, the 
calculation of these coupling matrices is numerically significantly more 
involved than in our previous approach. 

As in~\cite{RD}, splines are used to interpolate the 
various matrix elements between time steps. The time steps used to produce 
the splines are not uniformly distributed but carefully selected to take into
account the steepness of the time dependence of the couplings.
Close to the bounce we use very short time steps to produce the splines  
($\sim 10^{-6}$) while far away we can increase the step up to $0.2$ (in 
units of $L$). Furthermore, due to the complex time dependence of some of the 
couplings very close to the bounce, exact integration of the matrix elements 
when propagating the solutions through the bounce is necessary in order to 
obtain satisfactory accuracy for large KK masses as in 
Fig.~\ref{f:more}. In this way the bounce is taken into account as accurate 
as possible. This affects the speed of convergence of the solutions w.r.t. 
$n_{\rm max}$, leading to the behavior described below Fig.~\ref{f:more}. 

\begin{figure}[ht]
\begin{center}
\includegraphics[height=6.7cm]{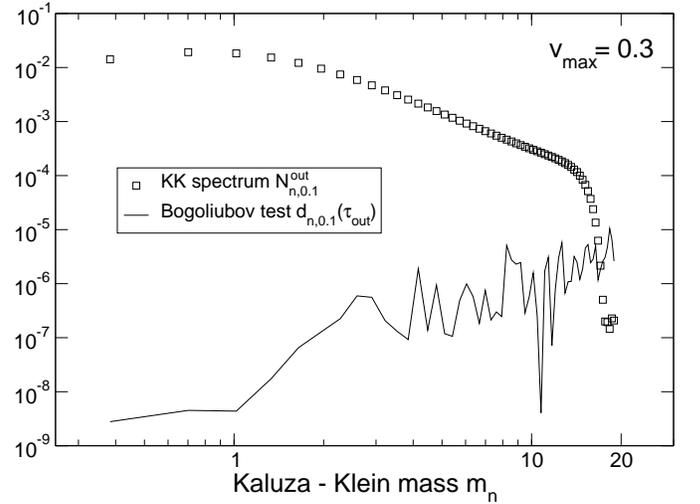}
\caption{Comparison of the final KK spectrum $N_{n,0.1}^{\rm out}$ 
and the corresponding quantity $d_{n,0.1}(\tau_{\rm out})$ given in Eq.~(D6) 
of~\cite{RD}. The quantity $d_{n,0.1}(\tau_{\rm out})$ is supposed to vanish 
identically, see~\cite{RD}. The comparison is a measure for
the accuracy of the $v_{\max}=0.3$ result depicted in Fig.~\ref{f:more}. In the 
region of the spectrum which is free from numerical artefacts, i.e. no 
dependence on cut-off $n_{\rm max}$, $d_{n,0.1}(\tau_{\rm out})$ is at least 
about two orders of magnitude smaller than the physically relevant quantity 
$N_{n,0.1}^{\rm out}$ indicating a satisfactory accuracy. 
\label{figure ap1}}
\end{center}
\end{figure}

As an indicator for the accuracy of our calculations 
we use the Bogoliubov test as described in Appendix D of \cite{RD} (Eq.~(D6)).
This is presented in Fig.~\ref{figure ap1} for the $v_{\max}=0.3$ result given 
in Fig.~\ref{f:more}. The quantity denoted by 'Bogoliubov test' and shown 
as solid line in  Fig.~\ref{figure ap1} should  ideally vanish. Given the 
complex nature of the numerical problem, the accuracy of the results is 
satisfactory for $m_n\lsim 10/L$.

\end{appendix}

\end{document}